\documentclass[aps,pra,reprint,superscriptaddress,nobibnotes]{revtex4-2}
\usepackage{graphicx}
\usepackage{dcolumn}
\usepackage{bm}
\usepackage{amsmath,amssymb,amsfonts}%
\usepackage{amsthm}%
\usepackage{physics}
\usepackage{multirow}%
\usepackage{booktabs}
\usepackage{xcolor}
\usepackage{lineno}
\usepackage{textcomp}
\usepackage{hyperref}
\usepackage[FIGTOPCAP,nooneline]{subfigure}
\hypersetup{%
 setpagesize=false,%
 bookmarks=true,%
 bookmarksdepth=tocdepth,%
 bookmarksnumbered=true,%
 colorlinks=true,%
 linkcolor=blue}

\begin{document}

\preprint{}

\title{Sub-Doppler cooling in a compact conical hollow mirror integrated with anti-Helmholtz coils}

\author{Shintaro Nagase}
\email[Contact author: ]{nagase@cns.s.u-tokyo.ac.jp}
\affiliation{Graduate School of Science, The University of Tokyo, 7-3-1 Hongo, Bunkyo-ku, Tokyo 113-0033, Japan}
\affiliation{National Metrology Institute of Japan, National Institute of Advanced Industrial Science and Technology, 1-1-1 Umezono, Tsukuba, Ibaraki 305-8563, Japan.}

\author{Naoya Ozawa}
\affiliation{Graduate School of Science, Kyoto University, Kita-Shirakawa-Oiwake-cho, Sakyo-ku, Kyoto 606-8502, Japan}

\author{Teruhito Nakashita}
\affiliation{Institute of Physics, Graduate School of Arts and Sciences, The University of Tokyo, 3-8-1 Komaba, Meguro-ku, Tokyo 153-8902, Japan}
\affiliation{Nishina Center for Accelerator-Based Science, RIKEN, 2-1 Hirosawa, Wako, Saitama 351-0198, Japan}

\author{Masaki Nakazawa}
\affiliation{Graduate School of Science, The University of Tokyo, 7-3-1 Hongo, Bunkyo-ku, Tokyo 113-0033, Japan}

\author{Kota Abe}
\affiliation{Department of Physics, Rikkyo University, 3-34-1 Nishi-ikebukuro, Toshima-ku, Tokyo 171-8501, Japan}

\author{Yu Nezu}
\affiliation{Department of Physics, Rikkyo University, 3-34-1 Nishi-ikebukuro, Toshima-ku, Tokyo 171-8501, Japan}

\author{Ryosuke Tsutsui}
\affiliation{Graduate School of Science, The University of Tokyo, 7-3-1 Hongo, Bunkyo-ku, Tokyo 113-0033, Japan}

\author{Hideki Ueno}
\affiliation{Nishina Center for Accelerator-Based Science, RIKEN, 2-1 Hirosawa, Wako, Saitama 351-0198, Japan}

\author{Hiromitsu Haba}
\affiliation{Nishina Center for Accelerator-Based Science, RIKEN, 2-1 Hirosawa, Wako, Saitama 351-0198, Japan}

\author{Yasuhiro Sakemi}
\email[Contact author: ]{sakemi@cns.s.u-tokyo.ac.jp}
\affiliation{Center for Nuclear Study, The University of Tokyo, 2-1 Hirosawa, Wako, Saitama 351-0198, Japan}

\date{\today}

\begin{abstract}
We demonstrate sub-Doppler cooling of rubidium ($^{87}$Rb) atoms using a compact, single-beam configuration based on a conical hollow mirror integrated with anti-Helmholtz coils.
The conical geometry converts a single incident beam into an effective six-beam magneto-optical trap (MOT) configuration while focusing the light towards the cone axis, thereby reducing the required laser power.
The number of atoms in the conical MOT is maximized at an incident intensity of 6.5 $\mathrm{mW/cm^2}$ in our conical MOT, several times lower than in conventional six-beam MOTs.
Under optimal conditions, we trap $\sim 1 \times 10^7$ $^{87}$Rb atoms.
We obtain sub-Doppler temperature of $T_r=13.4\pm 1.2$ \textmu K and $T_z=106\pm 11$ \textmu K by polarization gradient cooling (PGC), revealing a strong anisotropy that we attribute to the polarization pattern created by the conical mirror.
In addition, a modified release-and-recapture measurement in the MOT yields a radial temperature $T_r = 1.20\pm 0.02$ mK, approximately five times higher than that expected for comparable light-shift parameters in a six-beam MOT, in agreement with previous simulations.
Our results establish conical hollow mirrors as a practical and power-efficient platform for compact cold-atom sources and integrated quantum sensors.
\end{abstract}

\maketitle

\section{Introduction}
Laser-cooled atoms are key resources for a wide range of quantum sensing technologies, enabling ultra-sensitive measurement in atomic clocks \cite{RevModPhys.87.637,Ushijima2015,Nicholson2015}, atom interferometers \cite{RevModPhys.81.1051,PhysRevLett.78.2046,Rosi2014}, and searches for physics beyond the Standard Model \cite{RevModPhys.90.025008}.
The introduction of a hollow mirror or grating \cite{NOH2002269} for pre-stage cooling dramatically simplifies and miniaturizes the device, thus expanding the range of possible applications of laser-cooled atoms.
In addition to the device application, such compact cooling system are effective for spatially constrained experiments, such as laser cooling of heavy elements within fusion facilities \cite{10.1063/5.0118754}.
Single-beam magneto-optical traps (MOTs) including conical \cite{Kim:97,PhysRevA.64.013402}, pyramidal \cite{Lee:96}, tetrahedral \cite{Vangeleyn:09}, and grating \cite{Nshii2013,Vangeleyn:10,Lee:13} MOTs offer a promising route to compact and simple first-stage cooling.
In these systems, atoms are trapped by a single incident beam split into multiple beams by the mirror structure.
These systems have been integrated with atomic clocks \cite{Bowden2019,10.1063.5.0115382}, gravimeters \cite{Menoret2018,doi:10.1126/sciadv.aax0800}, gradiometers \cite{Stray2022,PhysRevA.105.022801}, and inertial sensors \cite{Wu:17,Lee2022}.

A conical MOT has unique properties arising from its axially symmetric intensity distribution.
The conical mirror surface focuses an incident beam towards the cone axis, creating a high intensity along the axis.
This focusing effect maximizes the number of trapped atoms with a significantly lower incident beam intensity \cite{PhysRevA.64.013402} compared with a typical six-beam MOT \cite{PhysRevA.65.015402}, thus realizing a power-efficient cooling scheme.
However, a numerical simulation has indicated that the radial atomic temperature in a conical MOT is much higher than that in the axial direction \cite{doi:10.1143/JPSJ.69.1366}.
Because the axially symmetric beam configuration in a conical MOT results in an effective scattering force that is approximately a central force, the Doppler-cooling effect is less effective for atoms moving on circular orbits around the cone axis.
Therefore, developing cooling methods that effectively reduce the radial temperature in this configuration is important.
Sub-Doppler cooling has been demonstrated in compact MOT configurations based on a diffraction grating \cite{Lee:13}, and more recently in back-to-back pyramidal and conical mirrors \cite{dkbj-8zfx}.

In this work, we demonstrate sub-Doppler cooling of $^{87}$Rb atoms using a single-beam MOT based on a conical hollow mirror.
We designed and implemented a compact assembly in which the conical mirror and anti-Helmholtz coils are integrated into a single mechanical structure, substantially simplifying and miniaturizing the MOT.
We show that the number of trapped atoms is maximized at a significantly lower beam intensity than that in a conventional six-beam MOT, consistent with a earlier work for generating cold atomic beams with a conical mirror.
We achieved sub-Doppler temperature along the radial and axial directions of $T_r=13.4\pm 1.2$ \textmu K and $T_z=106\pm 11$ \textmu K, respectively, in the conical MOT after applying polarization gradient cooling (PGC).
The measured radial temperature was $T_r=1.20\pm 0.02$ mK after MOT loading, approximately seven times higher than that in a conventional six-beam MOT with an equivalent light-shift parameter.

\section{Device concept of the conical MOT with compact mirror assembly}
\begin{figure}[t]
  \centering
  \includegraphics[width=0.5\columnwidth]{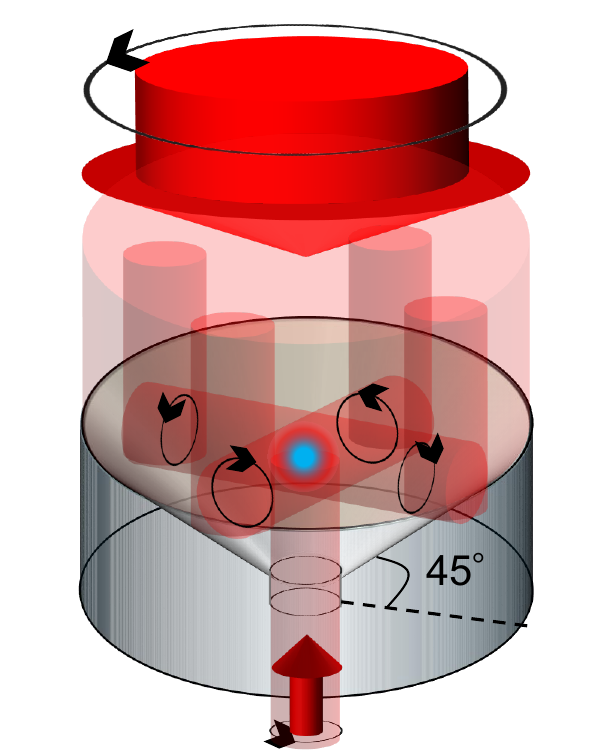}
  \caption{Beam configuration in a conical mirror.}
  \label{fig:conical_MOT}
\end{figure}

A conical mirror is formed by hollowing out a cylindrical substrate into a conical shape, with a hole at the apex along the cone axis as shown in Fig. \ref{fig:conical_MOT}.
The conical surface is tilted 45$^\circ$ relative to the bottom surface.
When a circularly polarized incident beam along the cone axis with a diameter comparable to that of the base is reflected on the conical surface, a polarization configuration along the radial direction appropriate for a MOT is generated in the hollow region.
For the axial direction, the portion of the incident beam transmitted through the hole at the cone apex passes through a quarter-wave plate (QWP) and is retro-reflected by a mirror.

\begin{figure*}[t]
  \centering
  \includegraphics[width=\textwidth]{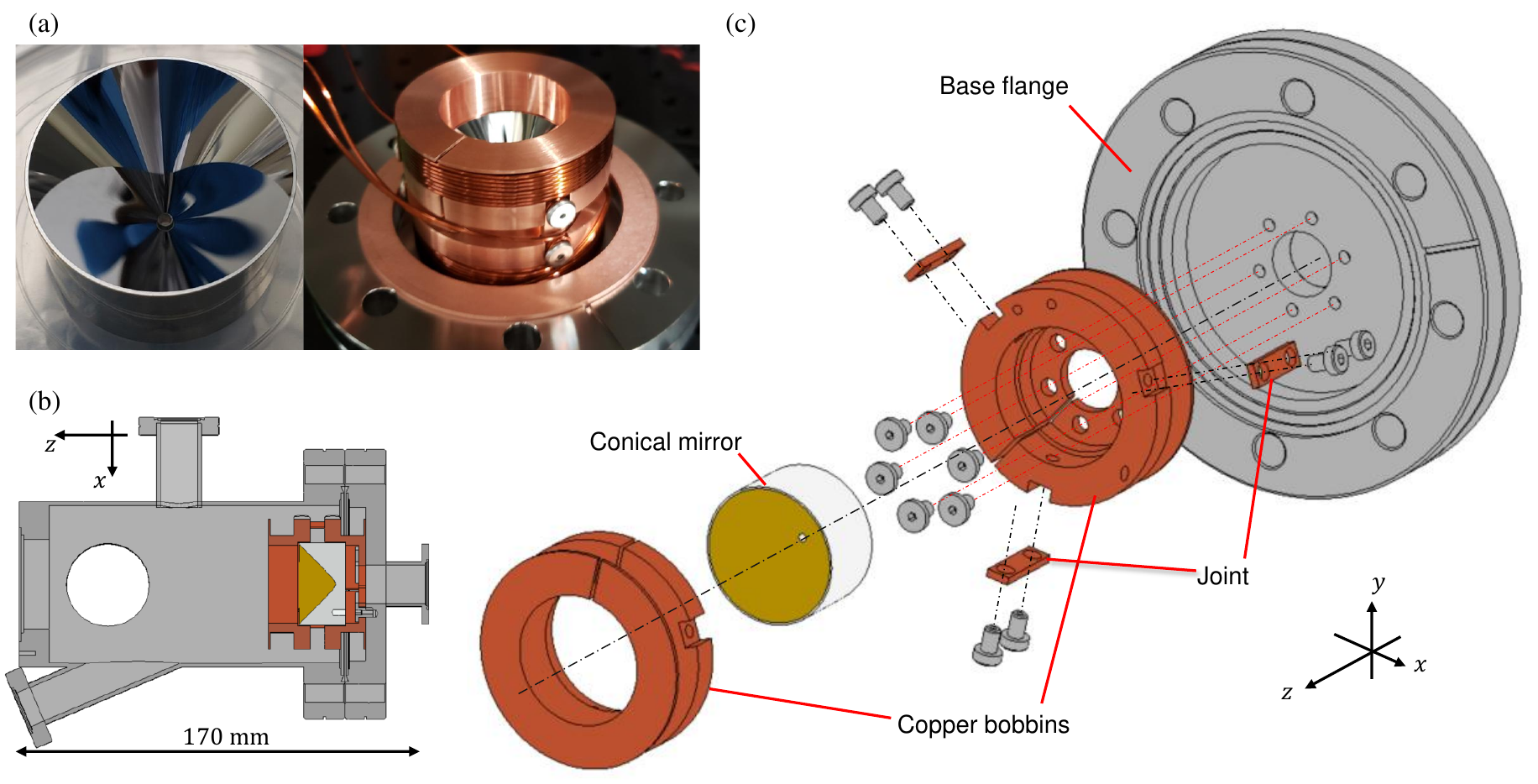}
  \caption{Design of the conical MOT apparatus integrated with anti-Helmholtz coils. (a) Photographs of our conical mirror and the conical MOT apparatus integrated with anti-Helmholtz coils. (b) Cross section of the main chamber. (c) Mechanical assembly of the conical MOT.}
  \label{fig:conical_mirror_assembly}
\end{figure*}

We fabricated the conical mirror using an aluminum cylinder with a base diameter of 35 mm and a height of 19.5 mm.
The inner surface was mirror-finished to a surface roughness of $R_a < 10$ nm and coated with aluminum.
A hole of 2 mm diameter was drilled at the cone apex along the cone axis.
The conical mirror was clamped between two copper bobbins, which contacted the top and bottom faces of the conical mirror as shown in Fig. \ref{fig:conical_mirror_assembly}.
The bobbins were attached to a base flange with M4 screws, forming a compact assembly and realizing differential pumping in the second cooling region.
The winding on each bobbin consists of 8 turns in the axial direction and 7 turns in the radial direction, producing the required quadrupole magnetic field.
The magnetic field gradients produced by 1 A of current in the anti-Helmholtz coils were calculated to be 9.7 G/cm/A along the axial direction and 4.8 G/cm/A in the radial direction.
In this compact coil configuration, resistive heating should be considered.
From a typical current of 1.8 A, total length of the copper lead of 34.6 m, and cross section of the lead of 1.04 $\mathrm{mm^2}$, the thermal dissipation is calculated to be 1.80 J/s.
Considering only the specific heat capacity of the copper lead, the rate of temperature increase is 29 mK/s, which is negligibly small.

\section{Experimental setup}
We used two home-built external cavity diode lasers (ECDLs) for the trapping and repumping in the conical MOT.
Each ECDL was frequency-stabilized to the D2 line of $^{87}$Rb atoms in a vapor cell using frequency modulation spectroscopy.
Each output was amplified by a tapered amplifier and transferred to the MOT region through polarization-maintaining fibers.
An optical switching was achieved using mechanical shutters and acousto-optical modulators, and the frequency of the trapping beam was controlled by a voltage-controlled oscillator.

\begin{figure}[t]
  \centering
  \includegraphics[width=\columnwidth]{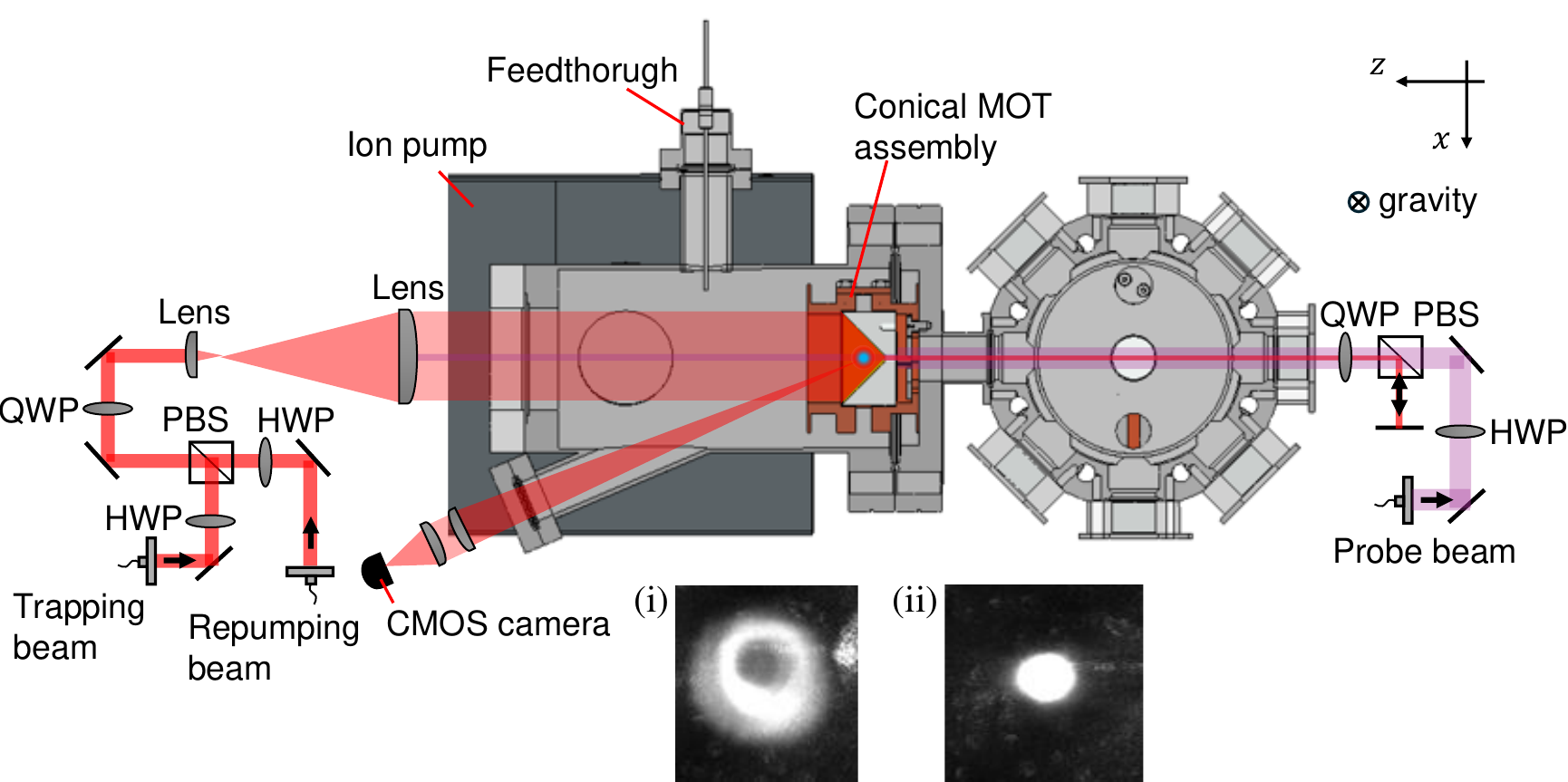}
  \caption{Optical system for the conical MOT. The red lines indicate the trapping and repumping as well as the atomic fluorescence collected by the CMOS camera, and the purple line indicates the probe beam. The current for the anti-Helmholtz coils is supplied via a feedthrough. (i) Ring-shaped atomic fluorescence image observed without magnetic-field compensation. (ii) Spherical atomic fluorescence image observed with magnetic-field compensation.}
  \label{fig:optical_system}
\end{figure}
A schematic of the optical system for the conical MOT is shown in Fig. \ref{fig:optical_system}.
The trapping and repumping beams were overlapped on a polarization beam splitter (PBS) and then passed through a QWP to form circular polarization.
The combined beam was expanded by a pair of lenses to a $1/e^2$ diameter of 28 mm.
After passing through an viewport, the beam was incident along the cone axis ($z$-axis) of the conical mirror, which was mounted inside the vacuum chamber.
The incident beam that passed through the hole at the cone apex was retro-reflected using a QWP, PBS, and mirror.
At this PBS, a probe beam resonant with the cooling transition ($F=2$ to $F=3$) was overlapped and directed towards the MOT region from the apex side of the conical mirror.
The probe beam had a power of 1 mW and a $1/e^2$ diameter of 12 mm, and was aligned so that its center passed through the hole of the conical mirror.
The typical incident powers of the trapping and repumping beams were 20 and 2 mW, respectively.
Fluorescence from the trapped atoms in the conical MOT was imaged on a CMOS camera positioned at an observation port, whose optical axis was tilted 22.9$^\circ$ with respect to the $z$-axis.

The main vacuum chamber, which houses the conical mirror assembly, was evacuated by an ion pump (PST-030CU).
A pressure below $1\times 10^{-7}$ Pa was achieved without atom loading, and the pressure during the experiment was typically $1\times 10^{-6}$ Pa.
The $^{87}$Rb atoms were provided from an ampoule.
On the right side of the conical MOT, an octagon chamber was mounted, which was reserved for a future second cooling region.
This region was differentially pumped to a pressure below $1\times 10^{-9}$ Pa.
The conical mirror acts as a differential pumping element between the first and second cooling regions, so that the conical MOT can serve as a compact cold-atom source feeding a downstream stage.
The conical MOT and second cooling regions are connected via the tube at the apex of the conical mirror whose radius and length are 1.0 mm and 3.5 mm, respectively.
Using the conductance of an orifice $C_0=\bar{v} \pi R^2/4$  ($R=1.0$ mm: radium of the orifice, $\bar{v}\simeq 270$ m/s: the average thermal velocity of atoms) and the transmission probability $W_5$ \cite{10.1116/1.569064}, the tube conductance is $C=C_0 W_5\simeq 0.0818$ L/s.
Assuming that the gas flow through the tube is the only source of the background gas in the second cooling region, the pressure ratio of the first and second cooling regions is written by $P_\mathrm{2nd}/P_\mathrm{1st} =C/(C+S) \simeq 4.5\times 10^{-4}$, where $S=180$ L/s is the pump speed of the second cooling region.
Based on the above conditions, we estimate the achievable pressure to be $P_\mathrm{2nd}\approx 10^{-9}$ Pa given that $P_\mathrm{1st}\approx 10^{-6}$ Pa and other outdassing sources are removed.

The flux of the cold-atom beam generated using a conical mirror has been well studied in a prior experiment \cite{PhysRevA.64.013402}.
They confirmed that a continuous flux of $2.2\times 10^{10}$ atoms/s, a longitudinal velocity of 4 to 6 m/s, and a divergence of 150 mrad were obtained.
Thus far, we have captured approximately $10^7$ atoms in the second cooling region by supplying the cold-atom beam from our conical MOT.
Therefore, we confirmed that our conical mirror functions as a device for generating a cold-atom beam.

Typical fluorescence images of atoms trapped in the conical MOT are shown in panels (i) and (ii) in Fig. \ref{fig:optical_system}.
As observed in the previous studies \cite{Kim:97}, small misalignments between the zero point of the quadrupole magnetic field and the cone axis have a significant impact on the spatial distribution of the trapped atoms.
When the magnetic-field center was misaligned with the cone axis, the atomic distribution in the conical MOT formed a ring shape, as shown in panel (i).
To compensate the magnetic field, we placed three orthogonal pairs of compensation coils surrounding the conical MOT.
After compensation of the magnetic field, we acquired the spherical-shaped atomic fluorescence image shown in panel (ii).
Regarding the long-term stability of this conical-mirror MOT system, we assess that it is not inferior to that of a conventional MOT.
Indeed, once careful initial alignment of the optical axis, atomic-fluorescence signals were repeatedly observed over a five-month experiment.

\section{Performance of the conical MOT}
The number of atoms in the conical MOT depends on the detuning, beam power, and magnetic field gradient \cite{PhysRevA.46.4082,PhysRevA.62.013404,PhysRevA.65.015402}.
We characterized the MOT performance as a function of these parameters by measuring the number of trapped atoms from the fluorescence signal.
Here, the atomic fluorescence was evaluated based on the emission induced by a probe beam after a 5-s MOT loading period, during which both the magnetic field gradient and cooling light were switched off.

\begin{figure}[t]
  \centering
  \includegraphics[width=\columnwidth]{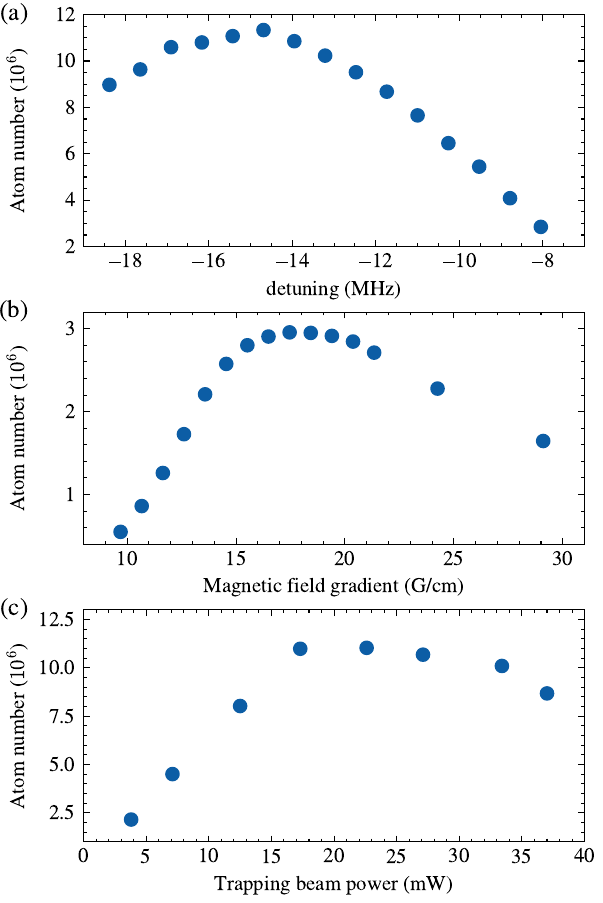}
  \caption{Dependence of the trapped number of atoms on (a) detuning $\delta$, (b) magnetic field gradient $dB/dz$, and (c) trapping beam power $P$. The other parameters were fixed at $dB/dz=17$ G/cm, $P=20$ mW; $\delta = -15$ MHz, $P=22$ mW; and $\delta = -15$ MHz, $dB/dz = 17$ G/cm, for panels (a), (b), and (c), respectively. The number of atoms was obtained from Gaussian fits to the atomic distribution, and the fitting errors are indicated as error bars. The error bars are smaller than the symbols.}
  \label{fig:MOTparameter_optimization}
\end{figure}
Figure \ref{fig:MOTparameter_optimization} shows the results of the parameter optimization.
We found that the number of atoms was maximized for a detuning of $-14.7$ MHz, a trapping beam power in the range of 17 to 22 mW, and a magnetic field gradient of 17 G/cm.
Here, the detuing is defined as the frequency difference between the trapping beam and the cooling transition of the $^{87}$Rb atom.
The dependence of the number of atoms on the field gradient and detuning was similar to that in a conventional six-beam MOT: the number of atoms increases with the gradient and detuning up to an optical value and then decreases for larger values.

The behavior as a function of trapping beam power is also similar: the number of atoms increases with power, reaches a maximum, and then slowly decreases at higher powers \cite{PhysRevA.62.013404}.
However, a notable difference was found in the value of the incident intensity giving the maximum number of atoms.
In our conical MOT, the optimal intensity was 6.5 $\mathrm{mW/cm^{2}}$, which was approximately four times lower than the optimal intensity ($\approx 25\ \mathrm{mW/cm^2}$ per beam) of conventional three-axis MOTs \cite{PhysRevA.65.015402}.
A similar trend appears in a prior study that generated cold-atom beams with a conical mirror, where the light intensity maximizing the atomic flux was reported to be approximately 80 times smaller than that in six-beam configurations \cite{PhysRevA.64.013402}.
This difference in the incident intensity is considered to arise from the focusing effect of the conical mirror surface, which concentrates the input light near the cone axis and enhances the local intensity in the trapping region.
We finally captured $1\times 10^7$ atoms in the conical MOT under optimal conditions.

\section{Realization of sub-Doppler cooling with the conical mirror assembly}
We measured the temperature of the atomic cloud after employing PGC using a time-of-flight (TOF) method \cite{PhysRevLett.55.48,PhysRevLett.61.169}.
Because the intensity distribution in the conical MOT is axially symmetric, we modeled the atomic distribution as an axisymmetric Gaussian with different temperatures along the axial ($z$) and radial ($r$) directions, denoted $T_z$ and $T_r$, respectively.

Based on the TOF measurement, the size of the atomic cloud along each direction after a release time $t$ is given by
\begin{equation} \label{eq:TOF_radius}
  \sigma_{z, r}(t) = \sqrt{\sigma_{z, r}(0)^2 + \frac{k_\mathrm{B} T_{z, r}}{m} t^2},
\end{equation}
where $\sigma_{z, r}(0)$ denotes the initial cloud size, $k_\mathrm{B}$ is Boltzmann's constant, and $m$ is the $^{87}$Rb atomic mass.
For each release time, we fit the fluorescence images with a 1D Gaussian on each axis and extracted the radii $\sigma_X$ and $\sigma_Y$ along the horizontal ($X$) and vertical ($Y$) axes of the CMOS camera oriented such that the $Y$-axis was parallel to gravity.
Because the imaging axis is tilted by $\theta = 22.9^\circ$ with respect to the cone axis, we converted the observed radii into the axial and radial radii.
Considering this geometric effect, we have $\sigma_r = \sigma_X$ and $\sigma_z = (\sigma_Y - \sigma_X \cos \theta)/\sin\theta$.
We extracted $T_r$ and $T_z$ by fitting $\sigma_z(t)$ and $\sigma_r(t)$ as functions of $t$ with Eq. (\ref{eq:TOF_radius}).

\begin{figure}[t]
  \centering
  \includegraphics[width=\columnwidth]{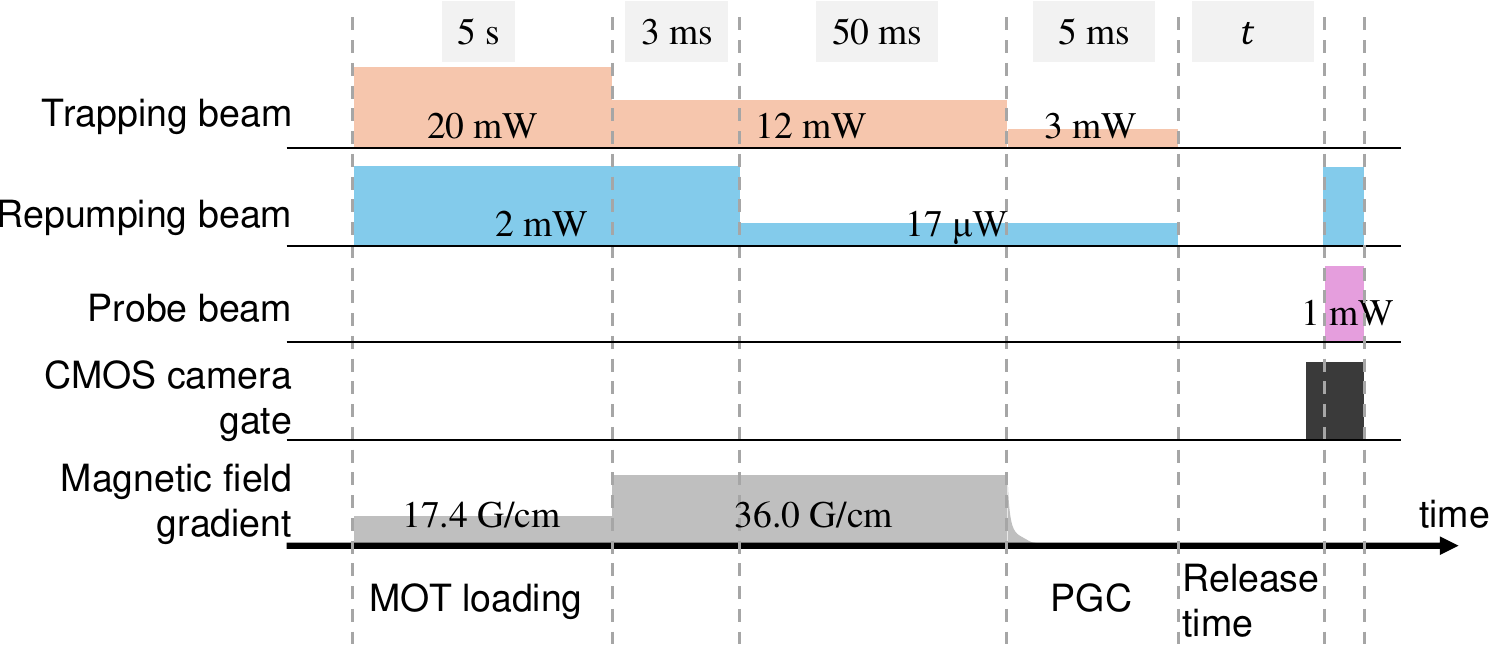}
  \caption{Timing sequence of the time-of-flight measurement.}
  \label{fig:sequence}
\end{figure}
To increase the atomic density before PGC, we employed a multi-step sequence.
The timing sequence for the temperature measurement is shown in Fig. \ref{fig:sequence}.
After a 5-s MOT loading, we compressed the trapped atoms by changing the detuning to $-33.1$ MHz, reducing the trapping beam power to 12 mW, and increasing the magnetic field gradient to 36 G/cm.
After 3 ms in this compression, we reduced the repumping beam power to 17 \textmu W for 50 ms.
These steps increased the atomic density by approximately a factor of two.
Subsequently, we applied PGC for 5 ms by switching off the magnetic field and further changing the trapping beam parameters: the detuning was set to $-60$ MHz and the power to 3 mW.
At the end of the PGC step, all beams were turned off, and the atoms were allowed to expand freely for a TOF duration.
After this expansion, the probe beam was irradiated on the atoms and the atomic distribution was imaged for analysis of the radius of the atomic cloud.

\begin{figure}[t]
  \centering
  \includegraphics[width=\columnwidth]{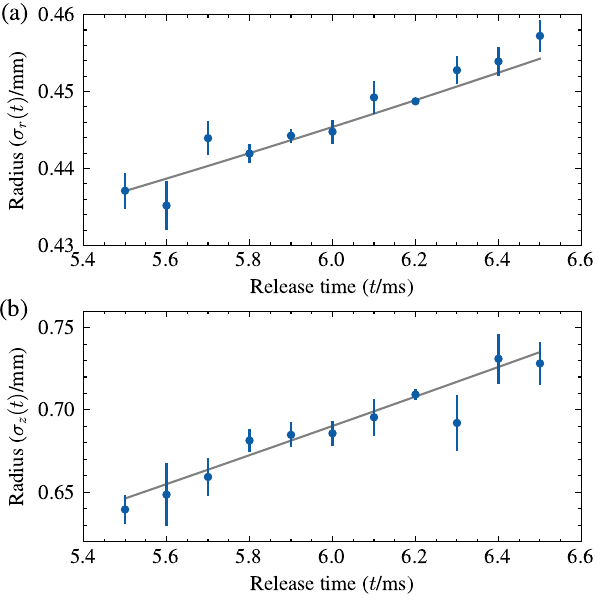}
  \caption{Atomic cloud radius as a function of the release time in the (a) radial and (b) axial directions. Each data point represents the average of three measurements, and the standard deviation is indicated by the error bars. The gray solid lines show fits by Eq. (\ref{eq:TOF_radius}).}
  \label{fig:TOF}
\end{figure}
The results of the TOF measurement are shown in Fig. \ref{fig:TOF}.
By fitting these data, the temperatures $T_z = 106 \pm 11$ \textmu K and $T_r = 13.4 \pm 1.2$ \textmu K were determined.
Both temperatures are below the Doppler limit for the $^{87}$Rb D2 transition ($T_\mathrm{D}=146$ \textmu K), demonstrating sub-Doppler cooling with our coil-integrated conical-mirror assembly.
These results reveals a strong anisotropy: the axial temperature is significantly higher than the radial temperature.
This anisotropy can be understood from the polarization distribution of the trapping light, which is crucial for PGC \cite{Dalibard:89}.
In our setup, the incident and reflected beams form a $\sigma^+$-$\sigma^-$ configuration.
In this configuration, the dominant sub-Doppler cooling mechanism arises from the atomic orientation induced as the motion of atoms in a spatially varying linear polarization field \cite{Dalibard:89}.
Near the cone axis, where atoms are captured, the focusing effect of the conical mirror makes the reflected light from the conical surface dominant.
If we consider only this reflected light, the polarization configuration is such that along the radial direction the linear polarization rotates in space, whereas along the axial direction the linear polarization angle remains nearly constant.
As a result, the conditions for PGC were more favorable in the radial direction, leading to a stronger cooling effect and a lower radial temperature, whereas the axial direction experiences weaker sub-Doppler cooling and retains a higher temperature.

\section{Evaluation of radial atomic temperature in the conical MOT}
We measured the radial temperature of atoms after the MOT loading.
In our setup, a standard TOF measurement is not suitable because simulations indicated that the atomic distribution in a conical MOT extends widely in the radial direction \cite{doi:10.1143/JPSJ.69.1366}, whereas our probe beam illuminates only atoms within a radius of $~ 1$ mm around the cone axis.
Under these conditions, the fluorescence image does not provide a reliable Gaussian profile from which to extract the cloud radius.

We thus employed a method similar to a release-and-recapture measurement \cite{RUSSELL2013313}.
The radial temperature was evaluated from the fraction of atoms that remained within the probe beam after a release time.
We assumed that the probe beam was infinite along the axial direction and that only the radial motion determined whether atoms remained within the illuminated region.
Let $r_c$ be the radius of the illuminated region and $v_r$ the radial velocity.
The fraction $f_r(t)$ of atoms still within the probe beam at time $t$ can be written as $f_r(t) = 2 \int_{0}^{v_{r_c}/v_T} ue^{-u^2} du$, where $u=v_r/v_T$.
We also defined $v_{r_c}=r_c/t$ as the critical radial velocity for atoms to remain within the probe beam, and $v_T=\sqrt{2 k_\mathrm{B} T_r/m}$ as the thermal velocity.
The integral yields
\begin{equation}\label{eq:RR}
  f_r(t) = A (1 - e^{-v_{r_c}^2 / v_T^2}),
\end{equation}
where $A$ is the normalization factor.
For a known radius $r_c$, we fit the measured fraction $f_r(t)$ as a function of the release time $t$ and extracted the radial temperature.
The effective radius $r_c$ was determined as follows.
For each TOF image, we summed the fluorescence signal along the $Y$-axis to obtain a 1D distribution along the radial direction.
We then performed a Gaussian fit to determine the center of the distribution and defined $r_c$ as the radius containing 95\% of the total intensity.
The average of $r_c$ over all TOF images was used in the analysis.

\begin{figure}[t]
  \centering
  \includegraphics[width=0.7\columnwidth]{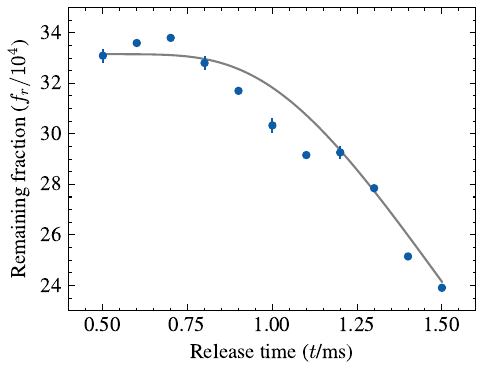}
  \caption{Remaining fraction as a function of the release time. Each data point represents the average of three measurements, and the standard deviation is indicated by the error bars. The gray solid lines show a fit by Eq. (\ref{eq:RR}).}
  \label{fig:RR_MOT}
\end{figure}
The measured fraction $f_r(t)$ is shown in Fig. \ref{fig:RR_MOT}, together with the fitting function.
From the fitting analysis, we obtained  $T_r = 1.20 \pm 0.02$ mK.
The radial temperature in the conical MOT seems higher than that in typical six-beam MOTs.
For a more quantitative comparison, we focus on the light-shift parameter $\Omega^2/(|\delta| \Gamma)$, which has a linear dependence on the temperature \cite{Wallace:94}, where $\Omega$ is the Rabi frequency, $|\delta|=2\pi \times 14.7$ MHz is detuning, and $\Gamma= 2\pi \times 6.1$ MHz is the natural linewidth. 
Because the total intensity needed for calculating the Rabi frequency  was difficult to measure directly in the conical MOT, we estimated our saturation parameter $s=I/I_s$ ($I$: total intensity and $I_s$: saturation intensity) from a previous study with sodium atoms in a six-beam MOT.
Based on the value reported in the previous study \cite{PhysRevA.65.015402} in which the MOT capture velocity was maximized, we set the saturation parameter to $s=8$, yielding our light-shift parameter $\Omega^2/(|\delta| \Gamma) \approx 1.65$.
At this light-shift parameter, the previous experiment \cite{Wallace:94} reported an atomic temperature of $\sim 160$ \textmu K in a six-beam MOT of $^{87}$Rb atoms, whereas the radial temperature in the conical MOT is approximately seven times higher.

This less effective cooling along the radial direction in the conical MOT is consistent with numerical simulations of atomic motion \cite{doi:10.1143/JPSJ.69.1366}, which indicated that the scattering force could be regarded as an effective central force.
Atoms moving on circular orbits around the cone axis experience cooling mainly through random changes in their trajectories arising from spontaneous emission.
As a result, the net cooling efficiency is reduced and the radial temperature is higher than that in conventional six-beam MOT geometries.
Our measurement provides an experimental confirmation of this behavior.

\section{Conclusion}
We demonstrated sub-Doppler cooling of $^{87}$Rb atoms using a conical hollow mirror.
We developed a compact conical MOT apparatus in which the conical mirror and anti-Helmholtz coils are integrated.
We characterized the dependence of the trapped number of atoms on the MOT parameters, and we successfully trapped $1\times 10^7$ atoms under optimal conditions.
By applying PGC, we reduced the axial and radial temperature to $T_z = 106\pm 11$ \textmu K and $T_r = 13.4 \pm 1.2$ \textmu K, respectively.
In addition, we measured the radial temperature in the conical MOT to be $T_r = 1.20 \pm 0.02$ mK, thereby experimentally confirming, in agreement with the previous simulations, that the temperature in the radial direction is relatively high in this geometry.

Our results establish that a single conical hollow mirror integrated with anti-Helmholtz coils is sufficient to generate large samples of cold atoms and to realize sub-Doppler cooling in a highly compact and simple geometry, while exhibiting characteristic anisotropic temperature and modified MOT dynamics.
The demonstrated platform is well suited as a front-end source for integrated cold-atom sensors and for compact cold-atom beam generation feeding second-MOTs or optical-lattice stages.

\section{Acknowledgments}
This work was supported by JSPS KAKENHI (Grant Nos. 19H05601 and 22K18273) and Grant-in-Aid for JSPS Fellows (Grant No. 22KJ1175).

\section{Data\ availability}
The data that support the findings of this article are not publicly available.
The data are available from the authors upon reasonable request.

\bibliography{apsrev.bib}

\end{document}